\documentclass[]{aastex631}

\def\ale{\mathrel{\hbox{\rlap{\hbox{\lower4pt\hbox{$\sim$}}}\hbox{$<$}}}}
\def\age{\mathrel{\hbox{\rlap{\hbox{\lower4pt\hbox{$\sim$}}}\hbox{$>$}}}}

\begin{document}

\title{Testing a Galactic Lensing Hypothesis with the Prompt Emission of GRB\,221009A}

\author[0000-0002-7777-216X]{Joshua S. Bloom}
\affiliation{Department of Astrophysics, University of California, Berkeley, CA 94720-3411, USA}
\affiliation{Lawrence Berkeley National Laboratory, 1 Cyclotron Road, MS 50B-4206, Berkeley, CA 94720, USA}



\begin{abstract}
Even at modest amplification, the optical depth to gravitational lensing through the Galaxy is $<10^{-5}$. However, the large apparent isotropic-equivalent energy of GRB\,221009A coupled with a path through low Galactic latitude suggests that the conditional probability that this particular GRB was lensed is greater than the very low {\it a priori} expectation. With the extreme brightness of the prompt emission, this Galactic lensing hypothesis can be constrained by autocorrelation analysis of {\it Fermi} photons on 0.1--1000\,ms timescales. In relating lensing mass, magnification, and autocorrelation timescale, I show that a lensed-induced autocorrelation signature by stellar lenses falls below the minimal variability timescale (MVT) expected from a black hole central engine. However, lensing by Galactic dark matter MACHOs ($M_l > 10-1000\,M_\odot$) could be confirmed with this approach. Regardless, at a peak $\gamma$-ray photon rate of $>30$\,ms$^{-1}$, GRB\,221009A represents a prime opportunity to measure the smallest MVTs of GRBs.
\end{abstract}



\section{Introduction}

At a redshift of $z_{\rm GRB}=0.1505$ \citep{2022GCN.32648....1D,2022GCN.32686....1C}, GRB\,221009A \citep{2022GCN.32632....1D,2022GCN.32636....1V} is among the lowest redshift long-soft GRBs known. With an isotropic-equivalent energy of $E_\gamma{(\rm iso}) > 3 \times 10^{54}$\,erg \citep{2022GCN.32668....1F} (the lower limit arising from detector effects) GRB\,221009A was one of the most (if not {\it the} most) intrinsically luminous GRBs ever observed (cf.~\citealt{2011ApJ...732...29C}). Observed in the direction of a low Galactic latitude ($b = 4.32^\circ,\,l = 52.96^\circ$), we expect larger stellar mass and dark matter columns than other well-observed GRBs far off the Galactic plane. This, coupled with (near) record-setting $E_\gamma({\rm iso})$ (and likely the lowest $V/V_{\rm max}$ of any GRB) suggests that the apparent prompt emission brightness could have been amplified by gravitational lensing.

Gravitational lensing of GRBs has been considered and purportedly observing previously in a cosmological context (\citealt{1986ApJ...308L..43P,1998ApJ...495..597L,2006ApJ...650..252H,2018PhRvD..98l3523J,2021NatAs...5..560P,2022ApJ...931....4L,2021ApJ...921L..30V}).  Here we (primarily) consider lensing by a Galactic object. In additional to amplification, lensing by a point mass forms two images of the source on the sky with an arrival time delay between each image \citep{2018PhRvD..98l3523J}:

\begin{equation}
\Delta t=\frac{4 G M_l}{c^3}\left(1+z_l\right)\left[\frac{\beta}{2} \sqrt{\beta^2+4}+\ln \left(\frac{\sqrt{\beta^2+4}+\beta}{\sqrt{\beta^2+4}-\beta}\right)\right],
\label{eq:t}
\end{equation}

\noindent with $\beta = \theta_s/\theta_E$, $M_l$ ($z_l$) the mass (redshift) of the lens, and $\theta_s$ the angular misalignment of the lens and the GRB. The Einstein radius is
\begin{equation}
\theta_E= \left(\frac{4 G M_l}{c^2} \frac{D_s - D_l}{D_s D_l}\right)^{1/2},
\end{equation}
\noindent with $D_s$ ($D_l$) as the angular diameter distance to the GRB (lens). On a cosmological scale, lensing by galaxies or clusters give rise to days-to-months delays between events or by seconds \citep{2006ApJ...650..252H} if the lens mass is an intermediate black hole or large dark matter halo ($M_l = 10^4-10^6 \,M_\odot$). Instead, as is our focus here, with $M_l = 30$\,$M_\odot$ in the Galaxy at $D_l = 8$\,kpc then $\theta_E=5.5$\,milliarcsec and $\Delta t = 1.2$\,ms at $\beta = 1$. These short timescales can only be probed with bright, fast changing GRBs. The total magnification $A(\beta) = (\beta^2 + 2)/(\beta \sqrt{\beta^2 + 4})$ establishes an implicit anti-correlation between $\Delta t$ and $A$.

The lensing time delay could give rise to an observable signature, a significant peak in the autocorrelation function of the $\gamma$-ray light curve \citep{2018PhRvD..98l3523J}, detectable only if the minimal variability timescale (MVT) \citep{2013MNRAS.432..857M} is less than $\Delta t$. As the observed variability in a GRB directly corresponds to the activity of the central engine in the internal shock model (cf.\ \citealt{2004RvMP...76.1143P}), the MVT lower limit should be set by the minimum possible timescale for variability in the central engine. Assuming emission from the innermost stable orbit (ISCO) of a non-rotating  $5\,M_\odot$ black hole, the MVT of GRB\,221009A must be greater than $(1+z_{\rm GRB}) R_{\rm ISCO}/c = 0.17$\,ms.

\begin{figure}[t]
\plotone{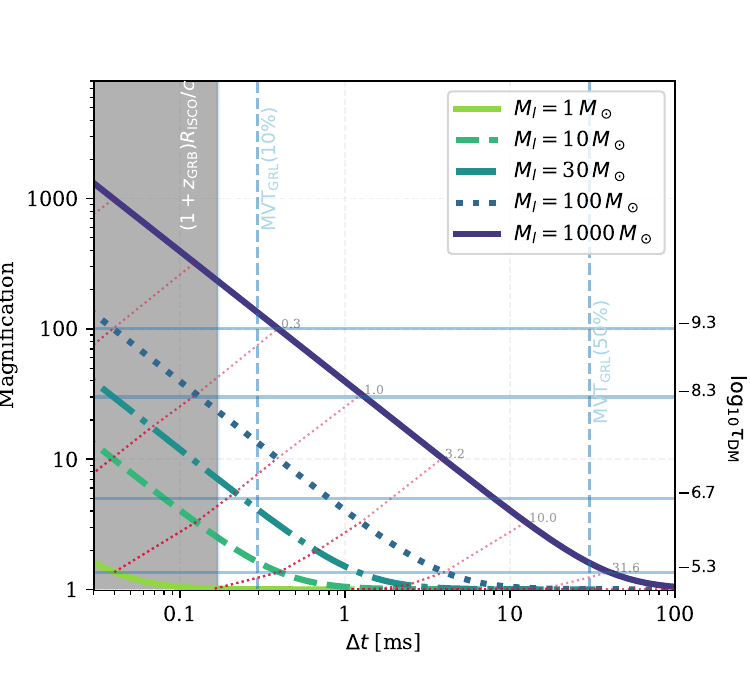}
\caption{The total magnification of GRB\,221009A as a function of correlation timescale $\Delta t$ for a range of Galactic lens masses at $D_l = 8$\,kpc. Lines of constant  $\theta_s$ are shown in dotted red, labeled in units of milliarcsec. The grey region corresponds to excluded times less than the redshifted timescale associated with a black hole central engine. The vertical lines correspond to the 10\% and 50\% minimal variability timescale of {\rm Fermi} events in the 300--1000\,keV range from \citet{2015ApJ...811...93G} (GRL), corrected to the redshift of GRB\,221009A using the median redshift ($z_{\rm GBL} = 1.72$) of the GRL sample.  The horizontal lines labelled at right show the approximate optical depth to massive compact dark matter objects in the Galaxy.
\label{fig:obs}}
\end{figure}

Figure \ref{fig:obs} demonstrates the phase space for lensing-induced autocorrelation  detectability for GRB\,221009A. Larger lensing masses, such as from MACHOs, could give rise to a detectable signal corresponding to large amplitude magnification. If the MVT of GRB\,221009A is at the 10th percentile of the GRL sample, then any autocorrelation lensing signature detected from a $M_l < 10 M_\odot$ would correspond to less than an factor of $2$ in magnification amplitude. 

There are potentially confounding matters to consider. First, as the ratio of the flux from the primary to the counter image location increases with decreasing $A$, small magnification events will also require larger signal-to-noise to be detected. Second, finite source effects limit the largest possible $A$. The largest radius at which internal shocks can dissipate energy before deceleration, is $R_\gamma \sim 4 \times 10^{13} (\Gamma_{\rm max}/100)^2 (T_{90}/10~{\rm s})\,{\rm cm} \approx 2.5 \times 10^{15} \rm{cm}$ with $\Gamma_{\rm max} \approx 440$ (following from \citealt{10.1046/j.1365-8711.1999.02970.x}). At an angular diameter distance $D_s = 546.9$\,Mpc the source size during the prompt emission was $R_\gamma/D_s \approx 3\,\mu$arcsec. Figure \ref{fig:obs} shows that finite-source effects are not relevant at $\Delta t \age (1+z_{\rm GRB}) R_{\rm ISCO}/c$. If we instead consider a lensing mass at cosmological distances, say $D_{s}/2$, then finite-source effects limit the largest amplitude to $A \ale 30$ for all $\Delta t$ with $M_l < 100 M_\odot$.

It is reasonable to ask what the probability that this event was lensed and, relatedly, what the chance that any GRB has been lensed with $\sim$ms time delays from a Galactic lens. At a fixed mass density and profile of dark matter in the Galaxy, the number density of primordial black holes \citep{2016PhRvL.116t1301B} and MACHOs \citep{2001ApJ...550L.169A, 2016PhRvL.117i1301M,2022A&A...664A.106B} decreases proportionally to the lens mass. The optical depth to lensing through the Galaxy can be calculated assuming a DM profile. I find $\tau_{\rm DM} = 4.8 \times 10^{-6} \beta^2$ in the direction of GRB\,221009A following from the calculation of \citet{1991ApJ...366..412G}. As a function of amplification this is $\tau_{\rm DM}(A) \approx 10^{-5} \times [A/(A^2 -1)^{1/2} - 1]$. Thus, the detectability phase space autocorrelation (say $A\sim[1.3-30]$) occurs with an optical depth from $5 \times 10^{-6}$ to $5 \times 10^{-9}$. In a cosmological context, \citet{2018PhRvD..98l3523J} found an optical depth of $\tau_{\rm DM, eg} \approx 0.15$ to the entire GRB-observable volume if all dark matter resides in $\sim30 M_\odot$ MACHOs. Given that GRB\,221009A originated from low redshift, the calculation for fast radio bursts (FRBs) is more relevant: \citet{2016PhRvL.116t1301B} found $\tau \approx 0.02$ to $z = 0.5$. Scaling by the ratio of the comoving volumes between $z=0.5$ and $z_{\rm GRB}$ gives an approximate extragalactic optical depth to this GRB of $\tau_{\rm DM, eg} \sim 0.0007$. Though the {\it a priori} expectation of lensing to any single event is evidently very low, the extreme apparent luminosity and low Galactic latitude clearly raises the conditional probability that GRB\,221009A itself was lensed. I do not attempt to calculate this conditional probability.

The {\it Fermi} GBM saw $0.375 \times 10^{6}$ photons per second during the peak of the prompt phase \citep{2022GCN.32642....1L}, corresponding to an average of 1 photon per 3 $\mu$s. As the MVT of GRBs is seen to decrease with increasing energy \citep{2015ApJ...811...93G}, focus on the higher energy photons should give access to the largest phase space for lensing autocorrelation discovery.  It is possible that intrinsic processes give rise to apparent  peaks in the autocorrelation function. However two observations would be the telltale confirmation of lensing for GRB\,221009A. First, the autocorrelation timescale should be the same at all energies. Second, over the very long duration of the $\gamma$-ray event ($>300$\,s; \citealt{2022GCN.32642....1L}), as the emission transitions from internal to external shocks, the MVT at high energy should increase; so long as MVT$(t) < \Delta t$, the measured autocorrelation peak should remain the same in the lensing scenario\footnote{In addition, VLBI measurements of the radio afterglow at later times could confirm lensing by revealing two images (suggested to me by W.~Lu, priv.~comm.). The separation would be $\sim 2 \beta \theta_E$ and the relative flux ratios and angular separation would place further constraints on $A$, $M_l$, and $D_l$.}. Dead time, pile up, and saturation effects will make an autocorrelation analysis on $0.1$--100\,ms timescales challenging (C.~Guidorzi, priv.\ comm.) but hopefully this note motivates such an effort.

%% Also note that the akcnowlodgment environment does not support long amounts of text. If you have a lot of people and institutions to acknowledge, do not use this command. Instead, create a new \section{Acknowledgments}.
\begin{acknowledgments}
I thank S.\ Bradley Cenko, Christiano Guidorzi, Raffaella Margutti, Liang Dai, Wenbin Lu, and Guy Nir for helpful conversations and comments.
\end{acknowledgments}



%% \software{astropy \citep{2022ApJ...935..167A}}



\newpage 
%\bibliography{refs}{}
%\bibliographystyle{aasjournal}



\end{document}